\documentclass[12pt]{article}

\usepackage{epsfig,bbm,bm,hyperref}
\input{epsf.sty}
\def\be{\begin{equation}}
\def\ee{\end{equation}}    
\def\baray{\begin{eqnarray}}
\def\earay{\end{eqnarray}}

\title{Structure calculation strategies for membrane proteins; a comparison study}

\author{Ileana Stoica\footnote{E-mail address: ileana@bri.nrc.ca}\\
National Research Council of Canada\\
 6100 Royalmount Ave.\\ 
Montr\'eal, QC, H4P 2R2, Canada}
\date{\today}
\begin{document}

\maketitle

\abstract{


Structure predictions of helical membrane
proteins have been designed to take advantage of the structural autonomy of
secondary structure elements, 
as postulated by the two-stage model of Engelman and Popot.
In this context, we investigate structure calculation strategies for
two membrane proteins with different functions,
sizes, aminoacid compositions, and topologies: the glycophorin A
homodimer (a paradigm for close inter-helical packing in membrane proteins) 
and aquaporin (a channel protein). 
Our structure calculations 
are based on two alternative folding schemes: a one-step 
simulated annealing from an extended chain conformation, and a
two-step procedure 
inspired by the grid-search methods traditionally used in membrane protein predictions.
In this framework, we investigate rationales for the utilization of
sparse NMR data
such as distance-based restraints and residual dipolar couplings in
structure calculations of helical membrane proteins.
}



\section{Introduction}

\hspace{12pt} A vastly used frame for understanding membrane protein
synthesis and folding is the two-stage model.

First introduced by Popot and Engelman (\cite{1}) and further elaborated by
White and Wimley (\cite{2}), the model proposes that the
stable conformation of a multi-spanning helical membrane protein is 
reached by interactions between pre-folded transmembrane (TM) helices,
without changes in the helical secondary structure. Typically, membrane protein 
structure predictions are "two-step" conformational searches and
follow conceptually the two-stage model (\cite{3,4,5}). The
assignment of individual transmembrane 
helices is the first step towards the prediction of a membrane protein
conformation (\cite{3,4,5}). It is usually considered that the success of
membrane protein prediction is 
largely determined by meeting the second challenge, of correctly
packing individually pre-folded helices (\cite{3,4,5}). 

In membrane
protein predictions, grid searches of the helical bundle
conformational space focus on helix-helix interactions and
assume a rigid canonical ${\alpha}$-helical backbone, in view of the two-stage
model of membrane protein folding 
(\cite{5,6,7,8}). 

Early conformational searches of the dimeric transmembrane region of
Glycophorin A from human erythrocytes have been designed to take
advantage of the secondary structure 
autonomy of individual helices (\cite{6,7}). These
and subsequent studies of other membrane proteins such as the
phospholamban homo-pentamer consisted of
multiple short {\it in vacuo} 
molecular dynamics simulations with different starting positions 
to ensure a comprehensive sampling of the helix-helix interaction space. 
Energy was used as a discriminating tool for selecting clusters of
most probable conformations (\cite{4}), while data
coming from mutational analysis, 
NOESY experiments, cross-linking studies, and CD-FTIR  were used as
an additional selection criterion, 
and to reduce the complexity of the search (\cite{4,6,7,8,9,10,11}).

The pioneering work on the glycophorin A (\cite{6,7}), as well as
more recent simulations on helical membrane proteins such as the
integrin homo-oligomer (\cite{12}), the Influenza virus A M2 proton
channel and the HIV vpu
(\cite{4}), and EmrE - the multi-drug 
transporter from E.coli (\cite{13}), are beautiful
demonstrations of the predictive power of the computational approach
in membrane protein structure 
determination. 

At the same time, an exhaustive search of the helical bundle
conformational space may become computationally expensive, especially
if symmetry considerations cannot be applied (the case of hetero-oligomeric proteins).
One way to overcome some of the difficulties related to conformational
sampling is to incorporate
topological restraints in the structure prediction/calculation
procedure (\cite{14}). 

Alternatively, constraints such as those obtained from
chemical cross-linking, fluorescence resonance energy transfer
(FRET), disulfide mapping, NMR, and site-directed spin labeling
combined with EPR can be incorporated in the design of penalty
functions that retrieve near-native conformations of helical bundles
from a collection of misfolded structures (\cite{15}). 
In these approaches, the
experimental distance restraints are used
in conjunction with structural rules derived from the analysis of
helix-helix interactions in known protein structures (\cite{4}).

The current work focuses on structure calculation strategies with
sparse NMR data. The most vastly used NMR restraints are
distance-based restraints which can be obtained from Nuclear
Overhauser Experiments (NOE), 
and backbone and sidechain dihedral angle restraints which can be
extracted from coupling data. More recently, global
alignment restraints in the form of 
residual dipolar couplings (RDC-s) have been proved to substantially
refine protein structures or even provide reliable predictions in the
presence of sparse NOE data  
(\cite{16}). 

It has been shown (\cite{17,18,19}) that protonated methyl groups offer
substantial structural information 
and thus significantly contribute to NMR protein structure
determination. With the advances in multidimensional NMR spectroscopy,
it has become possible to assign backbone 
and side chain resonances in highly deuterated, selectively
methyl-protonated proteins, some of them of high molecular weight
(\cite{17,18}). 
Resonances from selectively methyl-protonated hydrophobic residues can
be especially useful in membrane protein calculations, given that the
most important characteristic 
of transmembrane helices (which makes them readily detectable as
secondary structure elements) is their hydrophobicity (\cite{20}). 
Moreover, oligomerization and 
helix packing in membrane proteins usually involve hydrophobic motifs;
glycophorin A is
a paradigm for helical packing and dimerization mediated by such
a motif
(\cite{1,3,20,21,22}). 
	 
In the present study we explore structure calculation 
strategies for helical membrane proteins. Our purpose is to 
develop a rational approach for the utilization of 
a minimal data set of restraints that can be obtained by solution NMR. 
We aim to tailor our computations to
helical membrane proteins; therefore, 
we start by acknowledging the importance of single transmembrane helix
structure calculations. 
We address the issue of helical canonicity and investigate the role of NMR
data in accurately predicting folds of single membrane-spanning helices.
We then identify ways in which the analysis of
NMR data such as NOE-s and residual dipolar couplings (RDC-s) can offer insight 
into helix packing in membrane proteins. The third step is to
assemble the single-helix calculations and the information on inter-helical interactions 
into a 
whole-protein structure calculation. While doing that, we seek to take
advantage of some features of the grid-search approach that have proved
so powerful in 
membrane protein predictions. We use a two-step approach whereby
helix-helix packing is pursued in an independent rigid-body simulated
annealing refinement against 
inter-helical NOE-s, and then compare the results to a set of standard
 all-protein 
extended chain simulations.

\section{Methods}
\subsection{The proteins studied}

We start with structure calculations on glycophorin A.  Glycophorin A
(GpA) is the only homo-oligomerizing helical structure that has been
solved so far. 
The GpA homodimer has often been used as a model system for the study
of transmembrane helix structure and association (\cite{21,22}). 
The GpA dimer was first predicted computationally by Treutlein et al
(\cite{6}), using results from mutagenesis studies (\cite{10,11}) to narrow
the search. 
The prediction was later refined, using an improved global search
method (\cite{7}). The most recent structure of the dimeric
transmembrame domain of glycophorin A 
(GpA), in dodecyl phosphocholine micelles, was determined by NMR
spectroscopy by MacKenzie et al (\cite{23}), and provided excellent 
agreement with the computational predictions. Glycophorin A is a
largely studied example of close inter-helical packing in membrane proteins
(\cite{20}). Two largely preformed helical surfaces associate to form
the stable dimer, with a minimal loss of conformational entropy, as
postulated by the two-stage model (\cite{20}).

Channel proteins possess special features that have to be accommodated
by the two-stage model (\cite{1}): their partially polar backbones
interact to sequester away a transmembrane space from the lipid
environment. This requires that the unfavorable energy of creating a
water-filled channel from fully folded transmembrane helices is
compensated for by the energiy of oligomeric association (\cite{1}). 
The aquaporin family of proteins is involved in the permeation of
hydrophilic molecules across membranes. Modeling the human water
channel protein with the sequence-related glycerol facilitator coordinates (GlpF- \cite{24})
provided a first atomic model for aquaporin (\cite{25}). The atomic 
structure of bovine aquaporin 
was independently determined by X-ray analysis (\cite{26}). Along with
structural insights, molecular dynamics simulations by two different
groups 
(\cite{27,28}) have elucidated essential
aspects of water transport and selectivity in aquaporin. The water
channel protein is a tetramer, 
each monomer containing six tilted, bilayer-spanning helices 
and two non-spanning short helices - a functionally critical motif
(\cite{29}).

\subsection{Computing the NMR restraints}

\subsubsection{Dihedral angles, hydrogen bonds, and inter-helical distance restraints:} 
 
To generate canonical helices from extended conformations, we enforced
backbone (${\Phi, \Psi}$) dihedral angle assignments by placing small bounds around
the canonical values 
(-57${^\circ}$+/- 5${^\circ}$, -47${^\circ}$+/- 10${^\circ}$). $N_i-O_{i+4}$ hydrogen bonds were also added
to the restraint set, with a 2.99A target value and a 0.5A margin, to
avoid over-constraining 
(see \cite{30} for a statistics of hydrogen bond parameters). 

The initial NOE list comprised distances below 5.0A (amide-amide,
amide-methyl) or 6.5A (methyl-methyl) between all amide or methyl
positions that would be protonated 
in a deuterated, Ala, Val, Leu and Ile (${\delta1}$ only) - methyl protonated
sample (\cite{19}). 

The restraints were then assigned to one of three distance bins,
corresponding to strong, medium and weak NOE resonances,
respectively. 
Strong resonances (distances between 1.8A and 3.5A) were assigned a
3.0A target value with 1.2A and 1.0A lower and upper bounds,
respectively. 
Medium resonances (between 3.5A and 4.5A) were assigned a value of
4.0A, with margins of 1.2A. 
Weak resonances (between 4.5A and 6.0A) were assigned a 5.2A value, with more relaxed margins of 1.5A (lower bound) and 2.0A (upper bound).
All the assignments were performed non-redundantly (no pair of protons
appeared twice in the restraint list). 
Methyl groups were treated as pseudoatoms. Only HN(i) - HN(i+1) and
HN(i) - HN(i+2) distances were used for the amide proton list. 
No intra-residue resonances were considered. In the case of the
Glycophorin A dimer, inter-monomer resonances were ascribed
non-ambiguously 
(labeled according to the residue number and chain ID).

Subsequently, distance and residue- based criteria were used to
prune the tabulated restraint list to an experimentally realistic data
set. 
The distance-based selection criterion presented by Gardner and Kay,
and extracted from actual frequencies of observation for various
resonances 
in experimental spectra, was used as a reference (\cite{17,18}). Thus, sparser NOE sets were obtained 
by retaining only 2/3 of the ``weak'' resonances, or 2/3 of the ``medium'' resonances and 1/4 of the ``weak'' ones.
The
GpA inter-chain NOE-s involved the following 
residues: Leu 75, Ile 76, Val 80, Val 84, Thr 87, and  Ile 88. 
The number of inter-helical NOE-s was varied from 21 to 4 with the help of the selection criterion described above.
The
BMRB-deposited NOE data set collected on GpA samples by MacKenzie at
al (\cite{23}) includes 12 non-ambiguous 
inter-helical distances comprising a similar stretch of residues,
which have been identified by mutation analysis to be critical in helix-helix
packing and dimerization (\cite{10,11}).

In
aquaporin, sets of 348 and 115 distance restraints were designed using the selective
methyl labeling pattern, 
between a total of 8 helices, 10 residues or more each: the 6 membrane-spanning helices, and the two 
connecting helices that enter half-way through the membrane.
In structure calculations that focused on the TM helical bundle, 
the 115 NOE data set was further pruned to 92 resonances, corresponding to
inter-helical distances between the 6 TM segments only.

\subsubsection{Computation of RDC-s:}

\hspace{12pt}The residual dipolar coupling $D_{AB}$ between spin 1/2 nuclei A and B
(e.g. H and N) is given by 
(\cite{31,32}):

\begin{equation}
D_{AB} \approx -{\gamma_A \gamma_B} S <1/r_{AB}^3> D_a [ (3cos^2 \theta_{AB} -1)
  + 3/2 R sin^2 \theta_{AB} cos(2\Phi_{AB}) ]
\end{equation}

where $\gamma_A$ and $\gamma_B$ are the gyromagnetic ratios of the two nuclei, $r_{AB}$ is
the distance between the nuclei,  S is the generalized order parameter
of the dipolar vector 
and reflects averaging due to fast local dynamics, and ($\theta_{AB}, \Phi_{AB}$) are
the polar angles describing the orientation of the internuclear bond
vector AB in the alignment 
coordinate frame. $D_a$ and R are the axial and rhombic components of the
alignment tensor, respectively. In practice, the parameters needed by
the CNS (\cite{33}) and 
X-Plor (\cite{34}) SANI (susceptibility anisotropy) module are only $D_a$ and $R$.  

A Fortran program was written to compute residual dipolar couplings using as
alignment frame the principal axes of the molecular moment of inertia
tensor I; the program also 
computes the rhombicity R using the moment of inertia eigenvalues, for
a given value of $D_a$ (a $D_a$ of 10Hz was used throughout all of the
tests). For consistency, 
the sign of the calculated $D_{a(HN)}$ is in agreement with the negative sign of the $J_{HN}$ couplings.
The rhombicity R was calculated as:
 
\begin{equation} 
R = \frac{D_r}{D_a} = \frac{I_{xx} - I_{yy}}{I_{zz} - \frac{I_{xx} + I_{yy}}{2I_{zz}}}
\end{equation}

where $I_{\alpha \beta} (\alpha, \beta = x,y,z)$ are the eigenvalues of the moment of inertia tensor (ordered such as $I_{zz} > I_{xx} > I_{yy}$).
The set of H-N RDC-s generated with the in-house routine was then used
as initial guess for a Pales (\cite{35}) fit to the 3D
structure; 
to match the range of $D_{NH}$ values computed with $D_a$=10Hz, a bicelle
concentration of 2.5\% w/v was used for Pales. For consistency purposes,
the dipolar coupling tensor axis 
was then re-cast (also with Pales) into the new (Pales-computed)
alignment frame. Subsequently, during the TAD simulations, the
orientation of the alignment axes 
was allowed to float.

\subsection{The simulation protocols}

\subsubsection{Rigid-body/torsion angle annealing:}

For the two-step procedure, the TM
helices were first folded independently 
from fully randomized chains. For this, the CNS (Br\"unger et
al, \cite{33}) torsion angle dynamics anneal.inp file was used, with the standard
settings, and with a Cartesian cooling stage added to the TAD slow-cool.
The force constants on the NOE and dihedral angle restraints were kept
constant, at 150kcal/mol $A^2$ and 200 kcal/mol $deg^2$ respectively, 
while those on the residual dipolar couplings were ramped up from 0.03
to 0.6 kcal/mol $Hz^2$ (\cite{36}).
Canonical dihedral angles and hydrogen bond restraints were assigned
as described previously.

The Xplor-NIH package (\cite{34}) was used for the conjoined
rigid-body torsion angle simulations with canonical helices. The first of the two annealing schedules
described by Clore and Bewley (\cite{37})
was used with minimal modifications. The rigid groups were comprised
of the backbone atoms of the helical residues, while side chains were
given full torsional freedom. The only restraints used at this stage were the
inter-helical NOE-s.

The rigid body/torsion angle protocol started with a high-temperature
(3500K) internal dynamics run to promote convergence, followed by 10
cycles of dynamics with van der Waals repulsion turned on, and then by slow-cool from 3500K to
100K, in 25K-temperature steps (136 cycles), each of 1,000 steps, with
a fixed time step of 10fs.  
The force constants for the nonbonded interactions were: 4kcal/mol for
the van der Waals term, with a radius scale factor ramped
geometrically from 0.4 to 0.8, and 
4 kcal/mol $deg^2$ for the torsion angle energy. For the inter-monomer
NOE restraints, force constants were set to 20 kcal/mol $A^2$ during the
high-temperature stage and 
ramped from 100 to 450 kcal/mol $A^2$ during the cooling stage. Following
slow-cooling, 1000 steps of rigid-body minimization were performed,
with the final values of 
force constants from the cooling phase. 

For the simulations reported in Fig.7 and 8, a higher temperature of 5000K was used to 
improve sampling, and 5000 cooling steps were employed. Results are reported on 100 structure ensembles. 
Variations of the two-step rigid-body scheme included the
incorporation of attractive van der Waals and electrostatic
interactions to mimic the low-dielectric 
membrane environment. After the high-temperature search and the
slow-cooling stage with a purely repulsive van der Waals term, 5000
steps of constant-temperature (300K) 
molecular dynamics stage and a final minimization were carried out,
all including Lennard-Jones interactions and electrostatics. The
parameters used for the 
Cartesian molecular dynamics were similar to those used by Adams et
al (\cite{7,8}): a non-bonded cut-off of 12.5A; a switching function
was applied to the van der Waals terms between 10.5 and 11.5A, and a
dielectric constant of 2 
was used for the cell membrane. The time step was 2fs.

\subsubsection{Standard torsion angle dynamics (TAD) annealing:}

We compare the predictions of the two-step approach to
standard structure calculations from an extended conformation,
which we will refer to as 
"the one-step procedure". To carry out "one-step" all-protein
structure calculations we used the torsion angle dynamics
(\cite{38}) standard annealing protocol of 
CNS (\cite{33}). Most of the force constants were set to the
default parameters in the CNS anneal.inp module. The same canonical
dihedral angle assignments, hydrogen bonds, RDC-s and inter-helical NOE-s
were used as in the two-step rigid-body 
simulations. During the cooling stages, the force constants on the
NOE-s, dihedral angles and SANI were chosen to match those used in the
two-step procedure. 
For single-chain simulations, 2000 TAD cooling steps and 3000 Cartesian steps were used, 
while for whole-protein calculations, the number of slow-cool TAD steps was increased to 
10,000 to improve convergence.
All simulations
started from fully extended chains; 100 structure ensembles
were generated, out of which the
10 lowest energy structures were used to compute the average backbone accuracy. 
The ensemble accuracy is defined as the mean helical backbone RMSD
relative to the reference (PDB) coordinates, as calculated by MOLMOL (\cite{39}).

\section{Results and discussion}

\subsection{Single TM segment calculations; the role of RDC-s}

In a set of preliminary single-chain simulations, 
we concluded that the most efficient approach to single
TM helix calculations is the 
strong imposition of canonical dihedral angle restraints and hydrogen bonds, in
conjunction with RDC-s. This strategy eliminates the need
for intra-helical distances, thus considerably reducing the number of 
necessary NOE assignments.
The procedure yields a backbone RMSD around 1A per TM helix (20 -
30 residues), which becomes an implicit source of RMS deviations in two-step
conformational searches 
starting with independently folded TM segments. 
 We noticed that using RDC-s to enforce alignment of N-H
bonds about the helical axis 
generally helped "correct" the backbone dihedral angles towards their
"true" (native-structure) values, thus alleviating the deviations
arising from imposition of canonicity (Table 1).
This was certainly the case of the GpA dimer, which has in fact been
calculated in the original NMR structural determination (\cite{23}) with canonical DA-s. 

In
transporters and channels the presence of non-canonical helices is linked to the
existence of polar aminoacids with 
functional roles (\cite{40}).  
Thus, it is to be expected that for some membrane proteins the
imposition of canonical backbone dihedrals will result in helical
topologies with poor accuracies. 
Moreover, we typically detect a breakdown for refinement by
RDC-s, corresponding to a backbone RMSD threshold of
1-2A per TM helix (result not shown).
Consequently, helical conformations with poor accuracies are unlikely
to be improved by the inclusion of dipolar couplings, if no additional
restraints such as 
intra-chain NOE-s are used. As an example we show the aquaporin
TM helix 2 (Table 1), whose C-terminus significantly deviates from the expected
backbone dihedral angles.
 For instance, Val 69 and Ser 73 have ${\Phi}$-angles more than 60 degrees
 lower than the rest of the helix. The segment is however identified
 as a "certain" (score: 1.296) 
transmembrane ${\alpha}$- helix from sequence information by the TopPred (\cite{41}) predictor. 
As can be seen in Table 1, RDC-s fail to improve the structure of this helix.

\begin{table}
\begin{center}
\begin{tabular}{|c|c|c|c|} \hline
  \textbf{Protein} & \textbf{Number}  & \textbf{Ensemble}   &
  \textbf {Ensemble}  \\
 \textbf{and helix} & \textbf{of residues} & \textbf{accuracy (A)}   &
  \textbf {accuracy (A)}  \\
 \textbf{} & \textbf{in the helix} & \textbf{with RDC-s}  &
  \textbf {without RDC-s} \\

GpA monomer& 27 & 0.91 +/- 0.14 & 1.26 +/- 0.16\\
AQP1 TM helix 1 & 30 & 1.23 +/- 0.49 & 1.65 +/- 0.33\\
AQP1 TM helix 2 & 25 & 2.36 +/- 0.32 & 2.33 +/- 0.26\\
AQP1 TM helix 3 & 27 & 1.21 +/- 0.36 & 0.98 +/- 0.1\\
AQP1 TM helix 4 & 19 & 0.44 +/- 0.04 & 0.85 +/- 0.29\\
AQP1 TM helix 5 & 22 & 1.77 +/- 0.05 & 2.17 +/- 0.13\\
AQP1 TM helix 6 & 20 & 0.72 +/- 0.03 & 1.22 +/- 0.11\\\hline
\end{tabular}
\end{center}
\caption{Structure calculations of single transmembrane helices from
extended chains with canonical restraints (dihedral angles, hydrogen
bonds) and with or without 
dipolar couplings. The following coordinates were used as reference
for RMSD calculations: human glycophorin A (PDB code: 1AFO, structure
1 of the NMR ensemble of MacKenzie et al (\cite{23}, 
1997), and bovine aquaporin (1J4N, X-ray structure, 2.2A resolution,
Sui et al \cite{26}). RMSD-s are calculated on the 10 lowest energy
structures out of 100 
structure ensembles. }
\end{table}

To further highlight the role of RDC-s in single-chain structure
calculations, we examine the sensitivity of our simulated dipolar couplings
to backbone helical deviations from native conformation and/or from canonical topologies.
It is established experimentally that the
periodic variation of the magnitudes of 
dipolar couplings in the backbone of a protein as a function of
the residue number is linked to the periodicity inherent in regular
secondary structure elements 
(\cite{42}). Thus, periodicities in RDC-s known as "dipolar
waves" can be used to identify helices, deviations from ideal helical topologies, and
to orient helices relative 
to an external axis (\cite{42}). 
In fact, molecular fragment replacement methods have been designed
(\cite{16}) which identify the fold of protein fragments from sets
of dipolar couplings by searching a database (\cite{16}).
In structure calculation
studies, the agreement between dipolar coupling data and the molecular
structure is often assessed 
by the parameter (\cite{16,43}):

\begin{equation}
\chi = \sqrt { \sum{(D^{sim} - D^{ref})^2}}
\end{equation}

where the summation runs over the simulated set of DC-s,
and the reference values are typically the experimental ones (in our
case, calculated from the atomic coordinates).
We select two helices (one from each protein) to illustrate the RDC
sensitivity to deviations in helical topologies, as compared to the
reference (PDB) structure (Fig.1).

We
also include as an upper limit the ${\chi}$ values that would correspond to
extended chains (fully randomized 
conformations) of identical aminoacid sequences (Fig.1). As can be
seen in Fig.1, simulated RDC-s are sensitive to deviations of the
helical backbone from ``ideal'' topologies; a ${\chi}$ of 25Hz corresponds to a canonical helical
topology (helical backbone RMSD around 1A).

\subsection{Helix packing; the role of RDC-s and distance-based restraints}

\subsubsection{Inter-helical distances (contacts):}

An interesting question concerning inter-helix packing in
membrane proteins is whether TM helices pack preferentially
against sequence neighbors (\cite{4,44}). 

To critically evaluate our simulated inter-helical NOE restraints, we
analyzed this question for aquaporin, 
while also performing a
statistics over several transporter/channel proteins for
better comparison. 
With the particular labeling pattern used in our study, inter-helical
NOE-s involve predominantly the hydrophobic 
aminoacids. We therefore do not expect our inter-helical NOE pattern
to offer a perfect description of helix-helix interactions in protein channels for instance, where 
functionally important polar residues are often placed at the helical
interfaces (\cite{20, 40}). However, as can be seen in Fig.2, the methyl labeling pattern applied to aquaporin
does reveal the existence of a significant number of NOE-s between non-consecutive
pairs of TM helices.

Indeed, a small statistics of backbone inter-helical contacts over 6
transporter and channel proteins among which aquaporin (Fig.3) revealed that,
in fact, aquaporin has the highest fraction of contacts between
helices separated by 1 or more TM helices down the sequence, 
which is well captured by the NOE pattern presented in Fig.2. 

We also note that the incorporation of distance restraints involving the two
partially inserted helices of AQP1 has a significant impact on the
accuracy of the calculated structure (see below). This may be useful 
in modeling pairs of helices involved in extensive contacts in
multi-spanning helical proteins.

\subsubsection{Helix crossing angles:} 

In the early GpA conformational searches (\cite{6}), ${\Omega}$ = +/-
45${^\circ}$ helix crossing angles were chosen as the optimal starting angles, as
they allowed convergence to 
stable right-handed coils. However, subsequent studies showed that
stable configurations could be identified by extensive sampling of the
entire space defined by 
the helical rotational degrees. In cases more complex than the
glycophorin homodimer (e.g. bundles of many non-identical helices) it
may be more important to have a 
good initial guess for crossing angles between pairs of helices, as
exhaustive conformational searches become expensive.

In Fig.4 we use simulated RDC data for such an initial guess.
To extract crossing angles from the RDC data, we used Pales (\cite{35}) and our
RDC initial guess to compute the alignment frame for each of the two
helices individually, then 
computed the tilt of each helix with respect to the laboratory z-axis
from the alignment tensor eigenvectors. We thus obtained a crossing
angle for the homodimer.
By starting with increasingly distorted helical backbones (with
respect to the PDB structure), we estimated the degree of over- or
under-prediction of the helix 
crossing angle for "non-ideal" helical segments. The calculation is
fairly robust to non-canonicity issues; the estimated crossing angles
proved to be affected by the 
quality of the individual helices only to a limited extent (Fig.4). There
remains the redundancy associated with the parallel/anti-parallel
orientation of the helices, which 
can be resolved from inspection of the inter-helical NOE-s, and that
associated with
the handedness of the coil (the sign of ${\Omega}$), which as we show later does not bias
the results of the two-step method 
(see the Results of the two-step procedure).

\subsection{Assembling the intra- and inter-helical information; simulations from pre-folded helices}

As postulated by the two-stage model, the second stage of membrane protein folding is the packing of pre-formed trans-bilayer helices. 
To this end, we performed a simulated annealing search for identifying the optimal
relative orientation of 
pre-folded helices. Recognizing that a force field-guided grid search
which relies upon the convergence of different starting structures to
particular 
energy minima may suffer from poor convergence in the absence of
experimental data (\cite{4}), we used inter-helical NOE
restraints to aid the 
simulated annealing search. We note in the single TM segment section
that the folding of individual chains relies heavily on backbone
dihedral angles, 
hydrogen bonds, and RDC-s. No intra-helical NOE-s are needed at
this stage ("stage one" in membrane protein folding). In stage two,
the only NMR 
data used to guide the search towards a stable conformation are
inter-helical NOE-s, while the helical backbones are kept rigid. 
In our tests, the
procedure implies a 
10-fold or more reduction in the number of resonances that can be used
in the structure determination, with the
methyl labeling pattern.

\subsubsection{Incorporation of attractive van der Waals and electrostatic energies:}

We examined the result of including a final stage, of molecular
dynamics with Lennard-Jones and electrostatic interactions, at the end of the conjoined
rigid-body/torsion angle simulated annealing. In this fashion we
attempted to improve accuracy by more realistically sampling helix-helix
packing interactions. Attractive van der Waals and electrostatic terms
were parametrized as described in 
the Methods section. When a
protein 
is modeled as rigid, 
the reorientation of internal groups does contribute to the
dielectric constant, unlike in free molecular dynamics, for instance,
where all reorientations of 
dipolar groups within the protein are included explicitly in the
simulation (\cite{45}). 
This observation makes the choice of the internal
dielectric constant in rigid-body simulations a tricky issue. 
In the present simulations, dielectric constants of 2 and 10 were used
to mimic the lipid environment. The differences in the final
ensemble-averaged accuracies were 
less than 0.2A. 
Torres et al also note that results are virtually indistinguishable for ${\epsilon}$=1 and ${\epsilon}$=2 (\cite{4}).
In Fig.5 are shown results obtained from initially parallel helices,
with varying degrees of backbone RMS deviation from the PDB
coordinates; a value of  ${\epsilon}$=2 (\cite{46}) was used for these calculations.

If helices were significantly distorted (more than 2A backbone RMSD per
helix), the improvement in ensemble accuracy upon inclusion of
electrostatic and attractive van der Waals forces was negligible (fell
within the
ensemble variance). On the other hand, the improvement was sizable for
``canonical'' helices (RMSD around 1A per helix).
In such cases, the positive effect of the final molecular dynamics round emphasizes
the role that inter-helical energies play, along with experimental
restraints, 
in guiding the search for stable conformations of ${\alpha}$-helical
bundles. We therefore chose to include the constant-temperature
molecular dynamics stage in all 
subsequent rigid-body conformational searches.

\subsubsection{Results of the two-step folding simulations:} 

In the first set of GpA simulations, we started with canonical helices
(generated as described in Methods), which were then
placed at 20-24A apart (measured between the ${\alpha}$-Carbons of Gly 83). We
varied the helix crossing angle ${\Omega}$, which in the case of a homodimer is double the
helix tilt angle, from -90${^\circ}$ to 90${^\circ}$ and covering both the right
(negative) and the left (positive)
 handedness.

The first main conclusion of our two-step structure calculations on
GpA is that the success of the rigid-body refinement is significantly
dependent on the quality of 
the predicted individual helices (Fig.5) and less dependent on the
helix-helix relative orientation in the starting structure
(Fig.6). Using the initial guess provided 
by the RDC data on the pair of canonical helices (crossing angles
around +/- 45${^\circ}$, see Fig.4) proved to offer a slight advantage in
accuracy (Fig.6).  
The problem of identifying the correct handedness of the dimer also
appears to be solved by the rigid-body procedure: while we do
occasionally see
low energy 
left-handed dimers, we note that the lowest energy ensemble has a majority of
right-handed helices, 
irrespective of the initial handedness (sign of ${\Omega}$). 
By non-redundantly re-numbering residues in the two GpA
chains 
(for NOE assignment purposes - see Methods), our RMSD calculation (as
well as visual inspection) includes the
handedness 
of the helix pairs, thus providing an implicit assessment the correct handedness.

As the initial crossing angle was varied between -90${^\circ}$ and +90${^\circ}$, a
difference of less than 1A in ensemble accuracies was spanned. 
The mean ensemble accuracies are about 1A better than the TAD ensemble accuracies (around 2A)
obtained by constraining ${\alpha}$-helical dihedral angles to canonical
values, and using 
the same 10 inter-helical NOE-s (Fig.7).

Finally, we report the performances of the two structure calculation
strategies (one and two-step) for our model proteins (Fig.7 and 8).

For GpA the two-step annealing of parallel prefolded helices is the best route towards an accurate global fold. 
The two-step procedure is also
computationally inexpensive, since only one independent single-chain
SA is needed in the first step to produce the starting canonical helices.
Increasing the number of inter-helical NOE-s by including more long-range restraints (up to 6.5A) improved the accuracy, 
while decreasing it dramatically reduced the quality of 
the calculated folds (Fig.7). With enforced canonical DA-s, variations in the number of intra-helical NOE-s 
(achieved by placing different ratios corresponding to frequencies of observation for medium and weak resonances, 
as described in the Methods section) had a negligible impact, falling within the ensemble variance. 
Different levels of DA imposition (rigid, strong, medium) had a modest impact on the helical fold, 
possibly due to the 
presence of hydrogen bonds which hold together the helical backbone. Overall, our simulations suggest that the 
assumption of canonical topologies for the monomers, together with the extraction of the maximum possible number of 
inter-helical distance restraints, is the best structure calculation strategy for GpA. In this context, 
the exploration of other labeling methods, including the use of paramagnetic spin probes for longer-range distances, 
may provide even better folds.

In the aquaporin
conjoined rigid-body/torsion angle simulations, the individual TM helices were folded separately from
extended 
chains as described previously, using RDC-s in addition to canonical
DA-s. The starting helices were picked from
the TAD ensembles 
whose average accuracies are listed in Table 1. 
Rigid-body refinement was then performed against the 92
inter-helical NOE-s, as described 
in the Methods section. For the corresponding one-step structure calculations from
an extended chain, the same inter-helical NOE-s were used as in the two-step procedure.
 
In aquaporin, the two-step procedure gave poor
accuracies (Fig.8). The size of the protein is definitely an issue, as
typically successful applications of the grid-search methods have been
reported for homo-oligomers (\cite{9,12,13}),
but not for bundles of 6
non-identical TM
helices.  
Other factors are the inter-helical
packing pattern (Fig.3) and the non-canonicity
of some of the aquaporin TM helices (Table 1). 
The average ensemble accuracy can be roughly approximated by the number of TM helices times the
average backbone RMSD per helix 
(with respect to canonical ${\alpha}$-helices of identical aminoacid
sequences).
For membrane proteins such as aquaporin, different labeling patterns and
longer-range 
distance constraints (such as those that can be obtained with
paramagnetic probes) will be needed to improve the quality of the fold with the two-step procedure. As suggested previously, 
RDC-s can also be used 
for a reliable initial guess of the crossing angles between pairs of TM helices.

The best folding strategy for the water channel was found to be the ``one-step'' procedure with canonical DA restraints, 
hydrogen bonds, and as many inter-helical NOE-s as available with our labeling pattern.
Fundamentally, the result is in agreement with the GpA conclusions: membrane proteins of different sizes, topologies and 
functionalities can be accurately folded by restraining helices into canonical formations
and focusing on the exhaustive assignment of inter-helical resonances (preferably longer-range). 
Adding intra-helical NOE-s does not bring visible improvements (Fig.8). 

Neither does changing the DA assignments to 
other target values:
A statistics over 160 transmembrane helices from 15 high-resolution
X-ray structures has provided an average value of 
(-60.7${^\circ}$+/- 11.7${^\circ}$, -44.7${^\circ}$ +/-
13.0${^\circ}$) for the helical $({\Phi, \Psi})$ angles in
channel proteins and solute transporters (\cite{47}).  
In the case of aquaporin we examined the impact of introducing
transporter/channel specific values in the dihedral angle assignments.
Tuning the target (${\Phi, \Psi}$) helical dihedral angles towards protein
class-average values did not produce an improvement in the
accuracy of the global fold (Fig.8). 

In the previous paragraphs we emphasized the similarity between the two test case proteins. 
While indeed both GpA and AQP1 calculations recommend the same strategy conceptually (canonical helices, 
exhaustive inter-helical assignments), the winning computational approaches are different. 
Our simulations suggest that GpA calculations benefit from the two-stage model assumption of secondary structure autonomy, 
while aquaporin helices need to be folded concertedly, technically speaking: intra- and
inter-helical forces need to be sampled together.
Moreover, the role of the two non-spanning helices is also highlighted in the AQP1 calculation, as their omission 
results in poor accuracies (Fig.8). This finding relates very well to
the observation of Sale et al (\cite{15}) concerning the application
of a scoring function comprising a variety of sparse distance
restraints to a set of six known membrane protein structures, among
which aquaporin. Aquaporin posed the biggest challenge to the approach,
which authors attribute to the removal of
contact penalties involving the two non-spanning helices (\cite{15}).

\section{Conclusions}

If canonical topologies are assumed for membrane-spanning helices as is the case
in most grid-search procedures (\cite{6,7,8,9,12,13}), 
the 
number of distance-based restraints that can be extracted from a typical solution NMR
spectrum can be significantly reduced to a subset containing mostly
inter-helical NOE-s. The assumption is that TM ${\alpha}$-helices can
be reliably
identified (by dipolar waves, hydropathy searches or chemical shift
assignment) and 
adequately constrained with a combination of dihedral
angles and hydrogen bonds. To compensate for an inaccurate TM helix
prediction or more importantly for deviations from canonicity, residual
dipolar couplings will be needed.

An average of 10 inter-helical NOE-s per pair of TM segments,
with the selective methyl labeling pattern considered in the present
study, resulted in medium resolution structures for GpA (1A - 2.5A accuracies). 
The procedure implies a significant (20-fold) reduction in the number of resonances necessary, 
thus eliminating the need for exhaustive intra-helical NOE assignment. Our results also recommend the use of 
other techniques such as spin labeling for the extraction of longer-range inter-helical distance restraints.

A canonical homodimer, GpA is best folded using the ``two-step'' method, inspired by the two-stage model of 
membrane synthesis and folding (\cite{1}).
The procedure benefits from a more realistic modeling of packing forces in a final short round of MD, as well as (slightly) 
from an initial guess of the helical crossing angle.

For a bundle of non-identical and non-canonical TM segments such as
the ``hourglass'' topology of aquaporin, the overall resolution of
the two-step method is low and scales 
almost linearly with the number of TM helices. The success of the ``one-step'' strategy suggests that in the water 
channel protein, unlike in GpA, 
helices need to be folded and packed concertedly.
Still, for aquaporin as for glycophorin A, the imposition of canonical helical topologies, combined with the exhaustive assignment of 
(long-range) inter-helical NOE-s, 
was found to be the best structure calculation 
strategy with sparse NMR data.

\section{Acknowledgments}
This work was started at the Univ. of Ottawa. The author wishes to
gratefully acknowledge Dr. Natalie Goto for valuable discussions 
and for an NMR spectroscopist's perspective.



\begin{figure}[htb]
\begin{center}
\includegraphics[width=0.6\textwidth,angle=0]{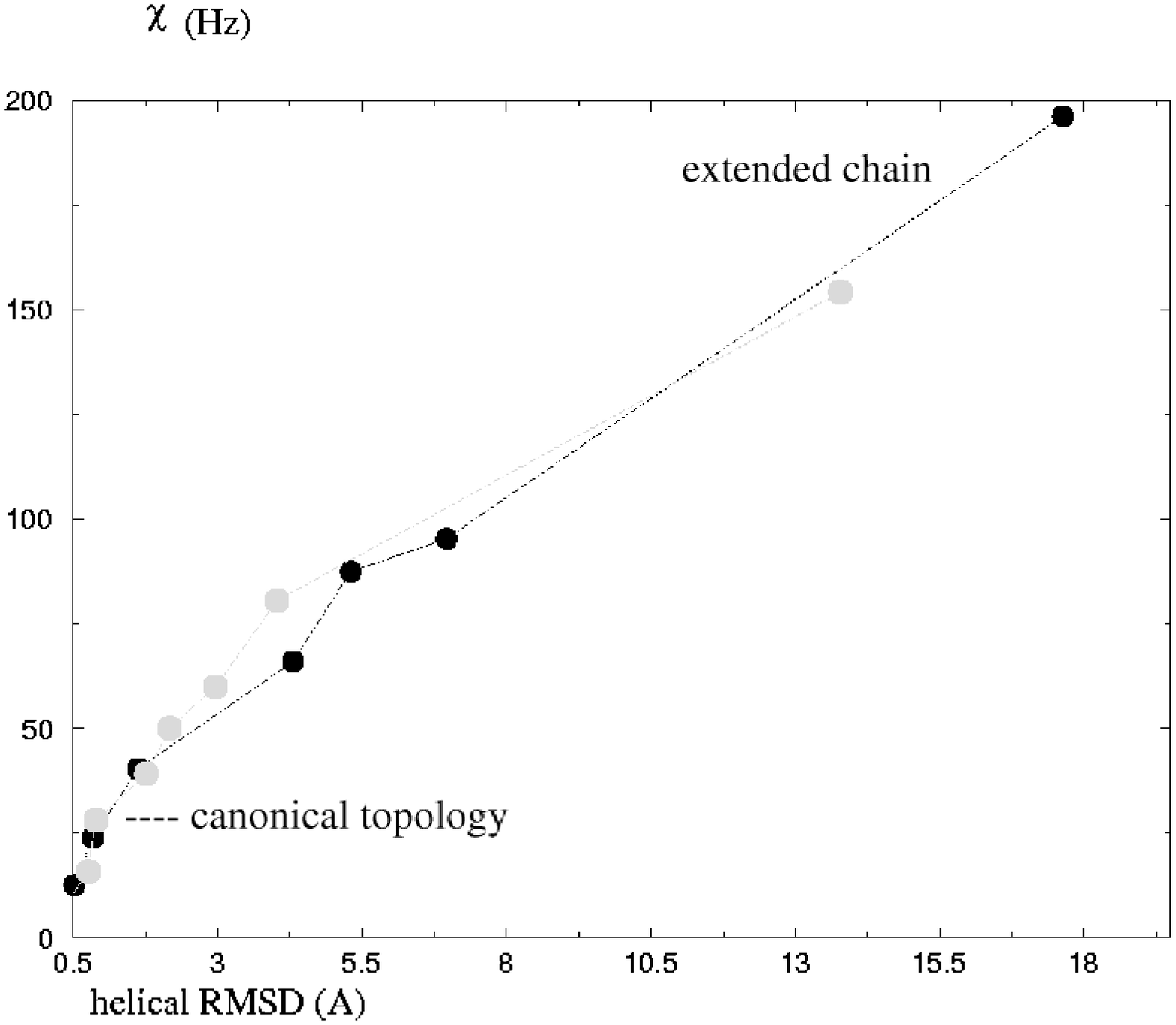}
\end{center}
\caption{Identification of helical regions from N-H RDC data. The
parameter ${\chi}$ is plotted for helices with increasing backbone
RMSD from the PDB structures. Black circles correspond to the GpA
monomer, and grey circles to AQP1 helix 1.
RMSD regions corresponding to canonical conformations
(generated with imposition of 
ideal ${\alpha}$-helical dihedral angles and hydrogen bonds), as well
as to extended chain conformations, are indicated on the graph.
}
\end{figure}

\begin{figure}[htb]
\begin{center}
\includegraphics[width=0.5\textwidth,angle=-90]{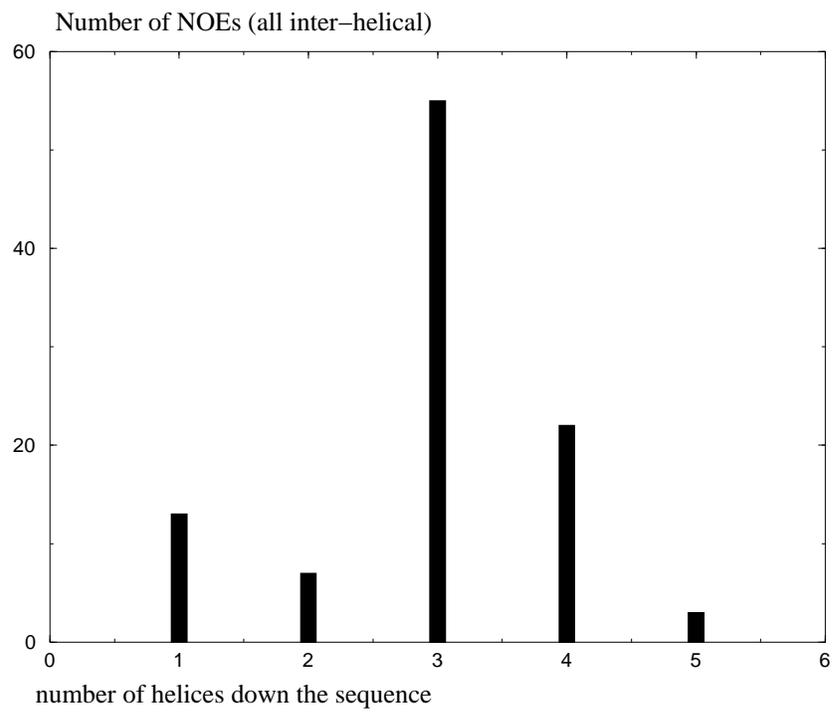}
\end{center}
\caption{Inter-helical NOE distribution in aquaporin (1J4N), with the
selective methyl labeling pattern, as a function of the TM helix consecutivity in the primary sequence. }
\end{figure}

\begin{figure}[htb]
\begin{center}
\includegraphics[width=0.7\textwidth,angle=0]{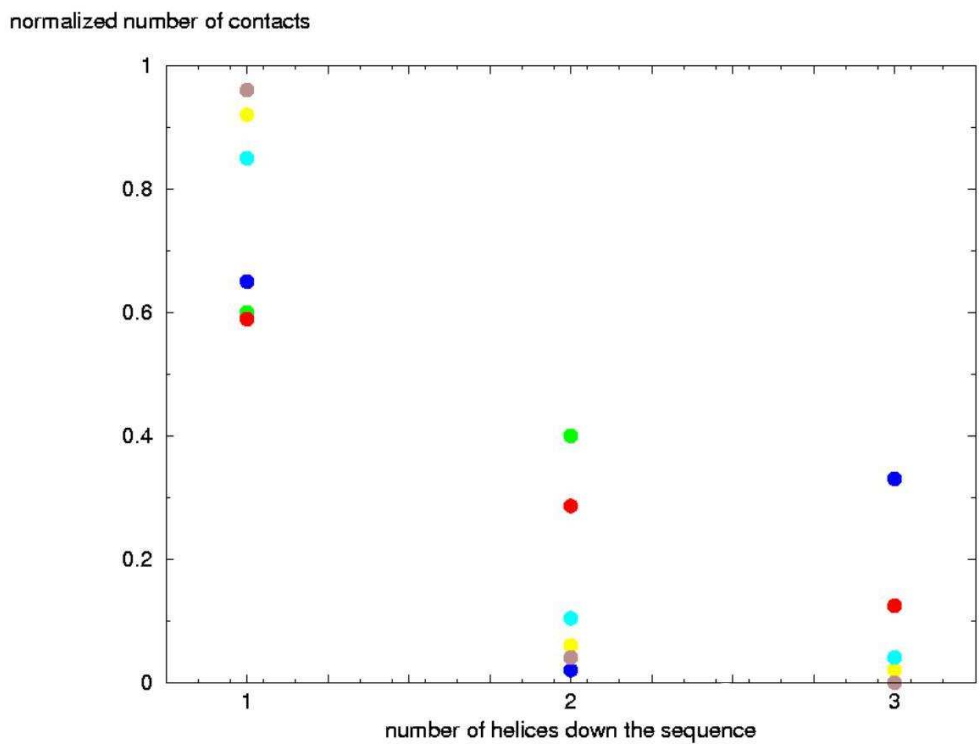}
\end{center}
\caption{Inter-helical backbone contacts in some transporter/channel
proteins: aquaporin (1J4N) in blue, the chloride channel (1KPK) in
red, the potassium channel (1BL8) 
in green, the lactose permease transporter (1PV7) in yellow, the
mechanosensitive channel (1MXM) in brown, and the vitamin B12
transporter (1L7V) in cyan. 
Backbone contacts between TM helices were defined with a cutoff of 6A (8A for the
mechanosensitive channel, which is very loosely packed), and counted
for: adjacent helices (n = 1), 
helices separated by one other TM helix (n = 2), and helices separated by two
other TM helices in the sequence (n = 3). Results are reported as fractions out
of the total number of 
contacts counted for a particular protein. A CNS (Br\"unger et al, \cite{33})
module was used to identify contacts within a given cutoff. 
}
\end{figure}

\begin{figure}[htb]
\begin{center}
\includegraphics[width=0.5\textwidth,angle=-90]{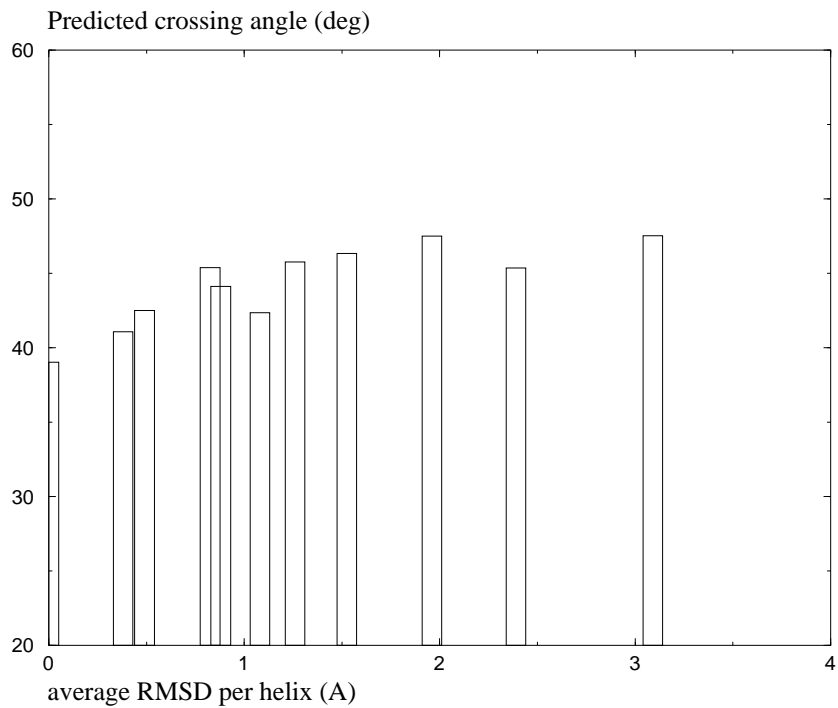}
\end{center}
\caption{Computation of helix crossing angles in the GpA homodimer,
from RDC data simulated for pairs of helices with varying degrees of RMS
deviation from the PDB coordinates. 
For each pair of helices, the crossing angle was calculated using the
alignment tensor eigenvectors produced by the Pales (\cite{35}) fit to $D_{HN}$ RDC
data sets computed as 
described in the text.}
\end{figure}

\begin{figure}[htb]
\begin{center}
\includegraphics[width=1.0\textwidth,angle=0]{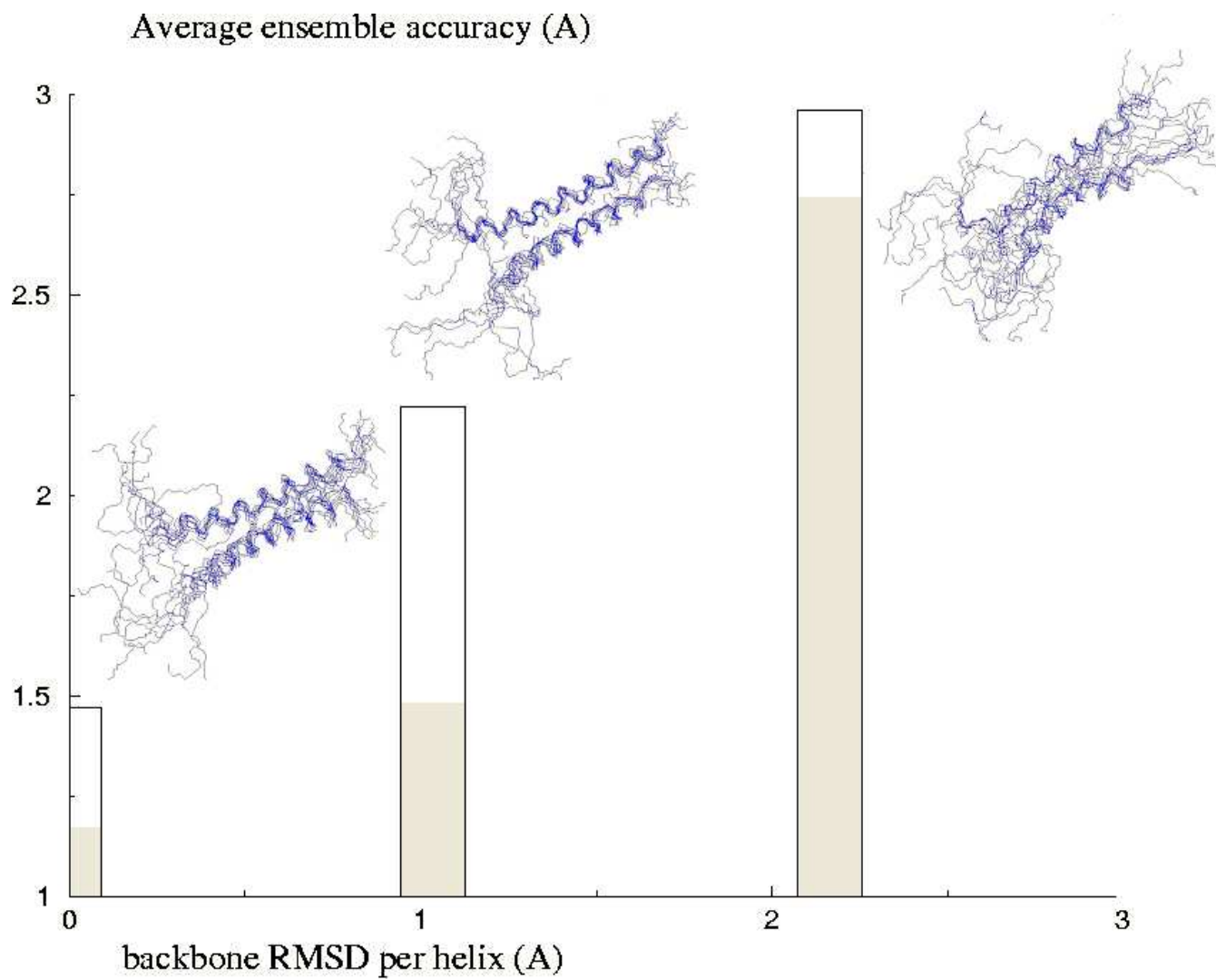}
\end{center}
\caption{Rigid-body/torsion angle annealing of GpA from pre-folded
parallel helices. Dependence of the ensemble accuracy (empty bars)
on the average backbone RMS deviation per helix in the starting
structure, relative to the reference (PDB) coordinates. Filled
bars: 
average ensemble 
accuracy of the lowest energy ensemble obtained upon
inclusion of a final constant-temperature rigid-body molecular
dynamics stage with attractive 
van der Waals and electrostatic energy terms. Lowest energy rigid-body
ensembles are 
also depicted above the corresponding accuracy/precision plots. The
structures correspond to the ``SA plus MD'' simulations (filled bars). }
\end{figure}

\begin{figure}[htb]
\begin{center}
\includegraphics[width=1.0\textwidth,angle=0]{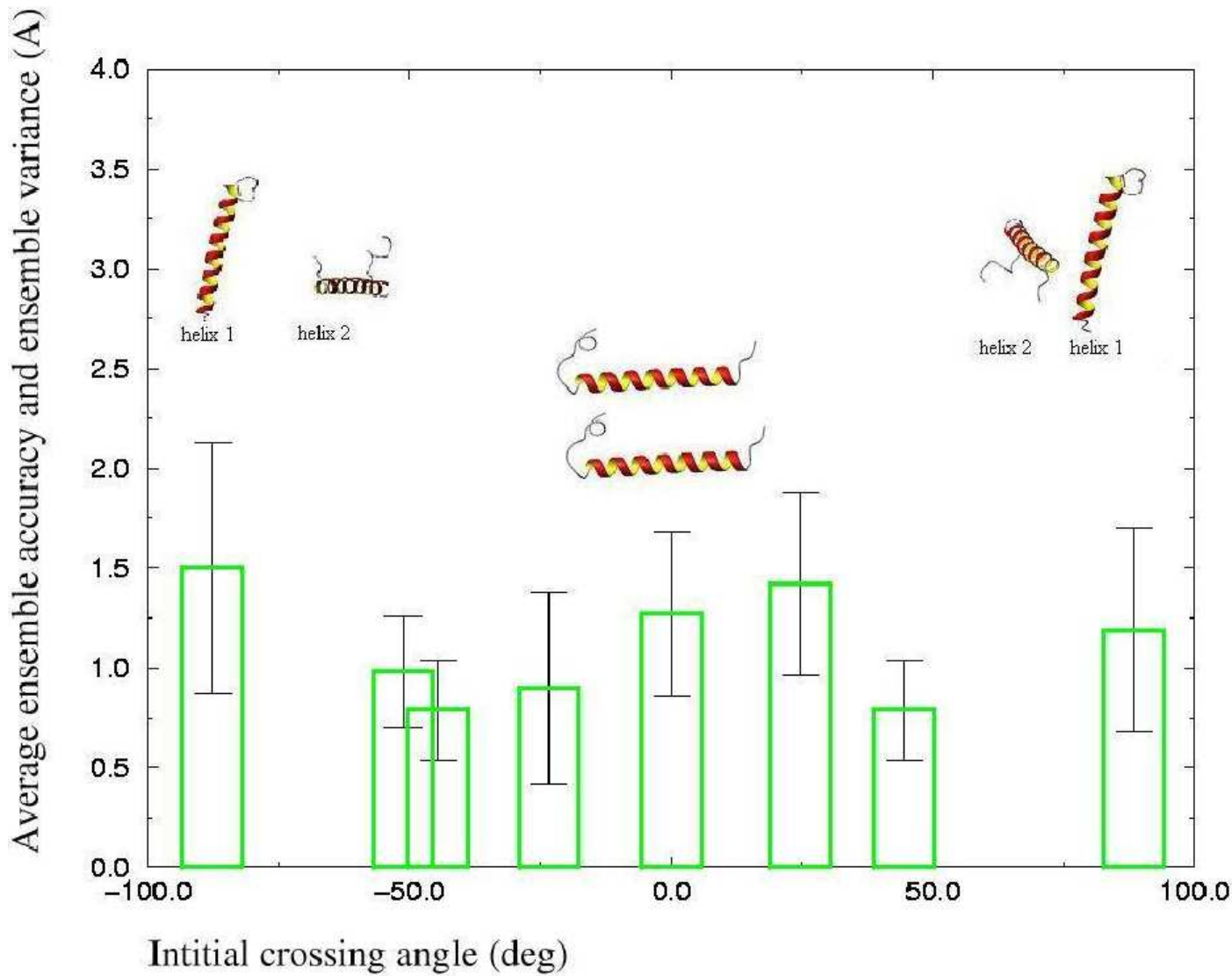}
\end{center}
\caption{Rigid-body refinement of GpA starting from pre-folded
canonical helices, placed initially at various crossing angles. Bars
represent average ensemble accuracies 
on 10 lowest energy structures out of 50 structure ensembles; the
variance of the ensemble accuracy is shown in thin lines. Attached on top are
the starting structures for 
the perpendicular (+/- 90${^\circ}$) as well as for the parallel helices. Helices are labeled unambiguously, 
as monomer 1 (left), and 2 (right). The
starting helices are obtained from TAD structure calculations with
canonical dihedral angles, 
hydrogen bonds and dipolar couplings, and have an average RMSD of
0.84A to the
PDB helices.}
\end{figure}

\vfill\eject
\begin{figure}[htb]
\begin{center}
\includegraphics[width=1.0\textwidth,angle=0]{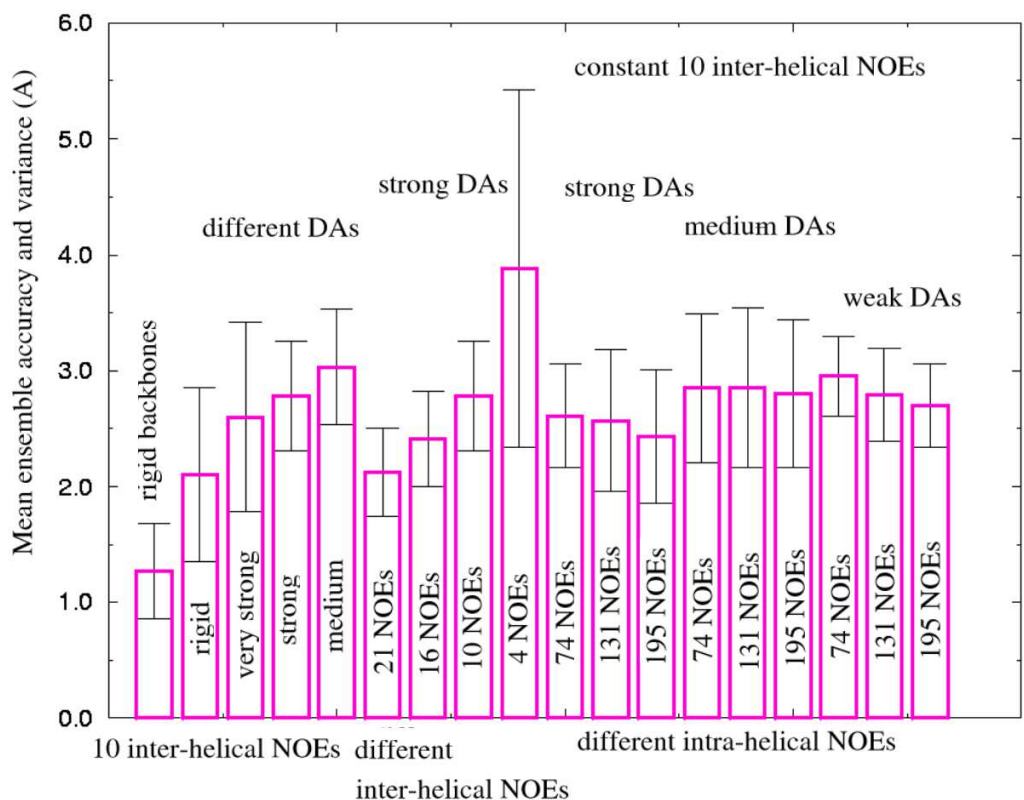}
\end{center}
\caption{Folding of glycophorin A}
The two methods used are:
Rigid-body/torsion angle slow-cooling from a collection of
canonical pre-folded TM helices using inter-helical NOE-s, and  
TAD-simulated annealing from an extended chain conformation with
dihedral angle restraints (DA-s), inter-helical NOE-s, and hydrogen bonds (HB-s).
With both methods, mean ensemble accuracy and ensemble variance are
reported on 10 lowest energy structures out of 100.
DA imposition is labeled as follows: rigid - (0${^\circ}$, 0${^\circ}$) bounds around canonical $\alpha$-helical dihedrals, 
very strong - (+/-2.5${^\circ}$, +/-5${^\circ}$) bounds, strong -  (+/-10${^\circ}$, +/-20${^\circ}$) bounds, medium - (+/-20${^\circ}$, +/-30${^\circ}$) bounds, and weak - (+/-40${^\circ}$, +/-60${^\circ}$) bounds.
\end{figure}

\vfill\eject
\begin{figure}[htb]
\begin{center}
\includegraphics[width=1.0\textwidth,angle=0]{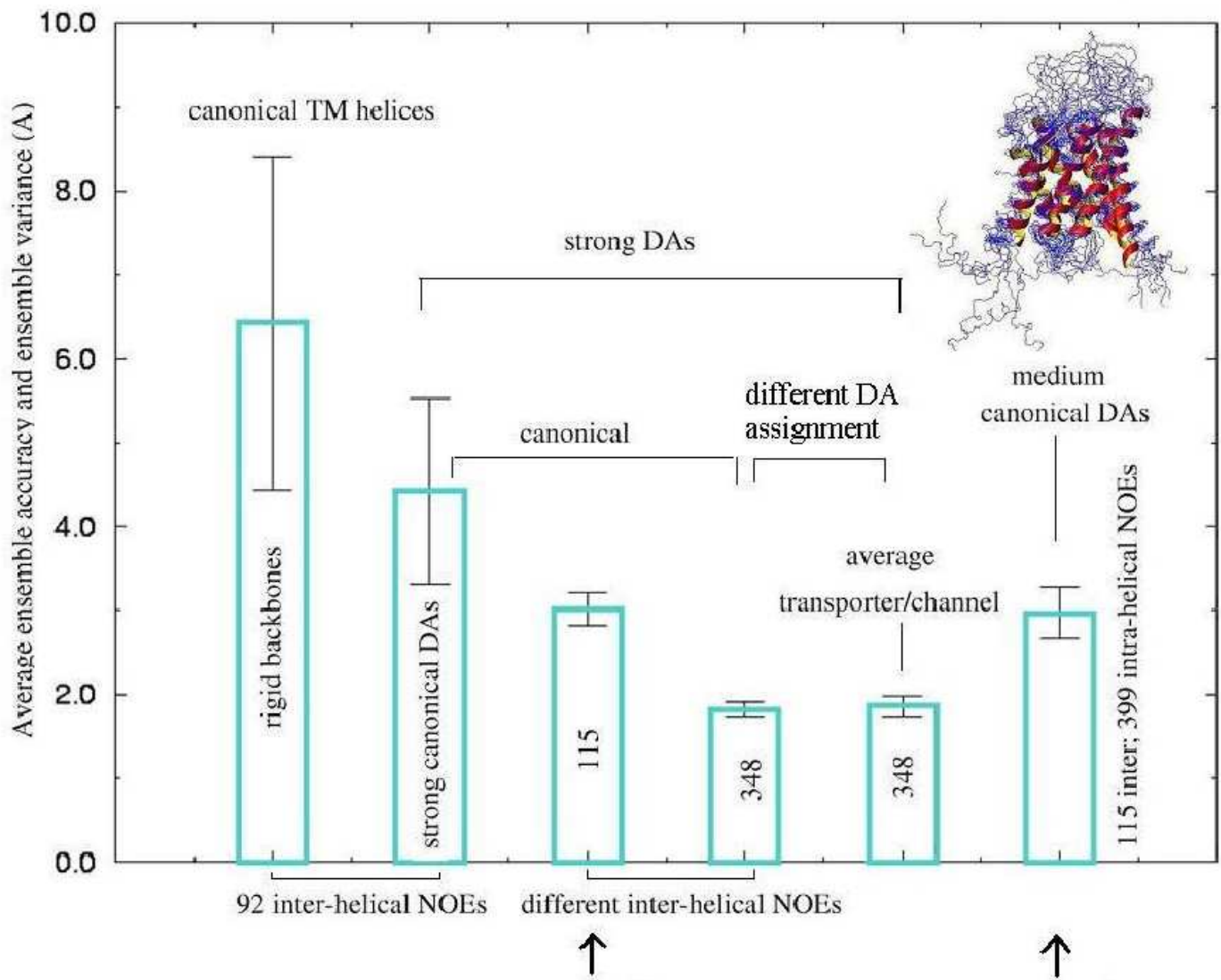}
\end{center}
\caption{Folding of the water channel} 
The two methods used are:
Rigid-body/torsion angle slow-cooling from a collection of
6 canonical pre-folded TM helices, with 92 inter-helical NOE-s, and  
TAD-simulated annealing from an extended chain conformation with
dihedral angle restraints (DA-s), inter-helical NOE-s, and hydrogen bonds (HB-s).
DA restraints are labeled as in Fig.7 (strong and medium). The target
values and bounds
for the transporter/channel-specific DA-s are given in the text.
The depicted ensemble corresponds to test 3 (strong canonical DA-s, 115 inter-helical NOE-s).
Accuracies are computed on backbone atoms from the 6 TM helices for
the first two tests, and for backbone atoms from 6 + 2
helices for the subsequent tests (including the 2 non-spanning
helices). 
Arrows identify structures calculated with the same inter-helical NOE-s and with or without intra-helical NOE-s.
\end{figure}
\end{document}